\documentclass{XrU2005}
\usepackage{graphicx}
\DeclareGraphicsExtensions{.pdf, .jpg}
\usepackage{subfigure}

\title{INTEGRAL observation of the X-ray burster KS 1741-293}
\author[1,2,3]{G. De Cesare}
\author[3]{A. Bazzano}
\author[4]{G. Stratta}
\author[3]{M. Del Santo}
\author[3]{A. Tarana}
\author[3]{P. Ubertini}
\affil[1]{Dipartimento di Astronomia, UniversitaÕ degli Studi di Bologna, Via
Ranzani 1, I40127 Bologna, Italy}
\affil[2]{Centre dÕEtude Spatiale des Rayonnements, CNRS/UPS, B.P. 4346, 31028
Toulouse Cedex 4, France}
\affil[3]{IASF-INAF, via Fosso del Cavaliere 100, I00133 Roma, Italy}
\affil[4]{OMP/LATT 14 Avenue E. Belin 31400 Toulouse, France}

\begin{document}

\keywords{X-rays binaries, INTEGRAL}

\maketitle

\begin{abstract}
KS 1741-293 was firstly detected  in 1989 with the X-ray wide field camera TTM (3-10 keV)  on board of the Rontgen-Kvant-Mir observatory. 
During these observations this source exhibited two X-ray bursts allowing to identify it  as a neutron star in a Low mass X-ray Binary. 
During the BeppoSAX/WFC monitoring of the Galactic Centre Region, KS 1741-293 was also reported  at a flux level of  6 mCrab in 
the 2-9 keV and 25 mCrab in the 9-25 keV energy range. Thanks to the deep and regular INTEGRAL observation of the Galactic Centre region, 
KS 1741-293 has been observed by the X-ray monitor JEM-X and the imager IBIS in a wide energy range, giving for the first time relevant information 
on its high energy behaviour. Furthermore,  two X-ray bursts have been detected by  JEM-X.
We report on  IBIS and JEM-X data analysis in terms of flux monitoring, spectral proprieties and bursts detection.  The data reduction has been done with the
most recent release of the standard analysis software (OSA 5.0).
\end{abstract}

\section{Introduction}
The first bursting sources were discovered already in 1975 with SAS 3 and OSO-8. All X-ray sources showing  type I burst  are low mass X-ray binaries (LMXBs). 
\cite{Woosley1975} and \cite{Maraschi1976} independently  discussed the origin of the phenomenon:  X-ray  bursts are explained 
by thermonuclear flashes of the material accreting from the companion star on the surface of the neutron star. 

The X-ray burster KS 1741-293 was firstly  reported  by \cite{Zand1991}  as one of the 2 new transient sources near 
the galactic centre during observation with the X-ray wide-field camera TTM on board the Rontgen-Kvant-Mir observatory, in 1989. 
The source was detected on 3 consecutive days and two  type I X-ray burst were detected  in the energy range 5.7$-$27.2 keV.  
The source was inside the error box of MXB1743-29, a bursting source detected in 1976 with SAS-3. More recently,  in the  
BeppoSAX era (1996-2002), this source was detected, together to a large sample of galactic sources, during the WFC monitoring 
of the Galactic Center Region \citep{Zand2004} at a peak flux of the order of 30 mCrab in the WFC energy band. KS 1741-293 was 
also detected by ASCA  during the 107 pointing observation on a 5x5 deg region around the Galactic Center showing an apparent variability 
by a factor of 50  while no burst have been reported by \cite{Sakano2002}. No hard X-ray detection has been reported by the first gamma-ray 
imager, SIGMA, and indeed the  source is not in the  hard X-Ray SIGMA survey, covering the 40-100 keV range \citep{Revnivstev2004}. 
KS1741-293 is listed in the BATSE/CGRO instrument deep sample as one of the 179 sources monitored along the CGRO operative life  
\citep{Harmon2004} even though it is not a ÒfirmÓ detection.

We show here for the first time the high energy spectrum of KS 1741-293 obtained with INTEGRAL. In section \S2 we report the INTEGRAL observations and data 
analysis, in section \S3 we briefly discuss the main results of this analysis. 

\section{Observation and data analysis}

\begin{table}
  \begin{center}
    \caption{KS1741-293 IBIS data set. The visibility periods 1, 2, 3 include all the public data and the core program data. 
    The last periods 4 and 5 include the core program data only. }
    \vspace{1em}
    \renewcommand{\arraystretch}{1.2}
    \begin{tabular}[h]{lrccc}
      \hline
      Period & Rev. & Start (MJD) &  End (MJD) & exp. (ks) \\
      \hline
      1  & 46-63     & 52698 & 52792 &306\\
      2  & 103-120 & 52871 & 52921 & 1333\\
      3  & 164-185 & 53052 & 53115  & 921\\
      4  & 229-249 & 53246 & 53306 & 159\\
      5  & 291-307 & 53431 & 53479 & 43\\
      \hline \\
      \end{tabular}
    \label{tab:table}
  \end{center}
\end{table}

The hard X- and gamma-ray observatory INTEGRAL was launched on October 17, 2002 by the a Russian PROTON launcher. The satellite makes revolution around 
the Earth in three days, along a highly eccentric orbit and the observing time is optimized by this choice. The wide-field Gamma-ray imaging and wide-band spectral 
capabilities of INTEGRAL coupled with the Core program strategy,  are a powerful tool  to further 
investigate the high energy behavior of X-ray bursters as firstly reported by \cite{Bazzano2004}.The scientific instruments on board are the hard X- and 
gamma-ray imager IBIS \citep{Ubertini2003} covering the energy band 20 keV$-$10 MeV,  the gamma-ray spectrometer SPI \citep{Vedrenne2003} in the same energy 
band but devoted to fine spectroscopy, the X-ray monitor JEM-X (3$-$35 keV) \citep{Lund2003}  and the optical camera OMC \citep{Mas2003}. We report 
here on the IBIS and JEM-X data analysis, performed with the last Off-line Science Analysis (OSA)  software (release 5.0).
Since IBIS is a coded mask telescope, the maximum of sensitivity corresponds to the fully coded field of view (FCFV) and for this analysis, we have selected the 
IBIS pointing in which KS 1741-293 is in the FCFV, including all the public data from  revolution 46 (2003-02-28) to  revolution 185 (2004-04-19), and 
the Core Program data \citep{Winkler2003}  from the revolution 46 to the revolution 307 (2005-05-19).   
The details of the observations are listed  in table \ref{tab:table}.  

\section{Results}

\begin{figure}
\centering
\includegraphics[width=0.7\linewidth,angle=-90]{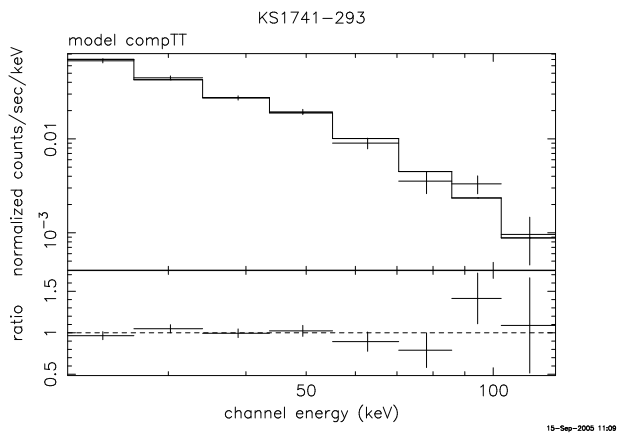}
\caption{IBIS/ISGRI KS 1741-293 spectrum obtained in the 3th visibility period  
\label{fig:spectrum}}
\end{figure}

We have monitored KS 1741-293 during a period of more then two year, from the end of February 2003 to end of May 2005, including in our analysis all the
public and core program data until April 2004 (periods 1, 2, 3). The source has been clearly detected in the periods 1 and 3. The signal in the period 2 is of order of 
10 sigma, but, a more refined analysis of the image has shown that this effect could be due to the 1A 1742-294 tail (21 arcmin far from KS 1741-293). The one second 
resolved light curves obtained from JEM-X  data in the revolution 53 (science window 58) and 63 (science window 92) exhibit two X- ray bust in the energy 
range 3Ð15 keV. The burst morphology  confirms previous observations with no double peaked time profile unlike to the profile 
reported for MXB1743-29 as discussed by  \cite{Zand1991}.

The spectra of LMXB bursters show, on average, similarities with the ones from Black Hole binary systems and consist of a soft, disk component with 
temperatures of the order of a few keV and a low energy gamma-ray tail. Sometimes a cut-off at around 50 kev is present in the spectra and 2 different states 
have been detected for some of them (for a comprehensive  review see \cite{Barret2001}).  The KS 1741-293 spectrum (fig. \ref{fig:spectrum}) has been extracted 
from the period 3 data. It is well fitted with a comptonized model (comptt) with an estimated plasma temperature of about 20 keV.

\end{document}